\documentclass[fleqn,usenatbib]{mnras}

\usepackage[T1]{fontenc}

\DeclareRobustCommand{\VAN}[3]{#2}
\let\VANthebibliography\thebibliography
\def\thebibliography{\DeclareRobustCommand{\VAN}[3]{##3}\VANthebibliography}

\usepackage{mystyle}

\usepackage{newtxtext,newtxmath}

\newcommand{\ngc}[1]{\ifthenelse{\equal{#1}{}}{NGC~4038/9}{NGC~#1}}

\makeglossaries

\newacronym{ad}{AD}{Anderson-Darling}
\newacronym{aca}{ACA}{Morita Atacama Compact Array}
\newacronym{alma}{ALMA}{Atacama Large Millimeter/Submillimeter Array}
\newacronym{co}{CO}{carbon monoxide}
\newacronym{casa}{\textsc{casa}}{Common Astronomy Software Applications}
\newacronym{carta}{\textsc{carta}}{Cube Analysis and Rendering Tool for Astronomy}
\newacronym{cn}{CN}{the cyanide radical}
\newacronym{dec}{Dec.}{declination}
\newacronym{fdr}{FDR}{false discovery rate}
\newacronym{fov}{FoV}{field of view}
\newacronym[firstplural=full widths at half maximum (FWHMs)]{fwhm}{FWHM}{full width at half maximum}
\newacronym{ir}{IR}{infrared}
\newacronym{ism}{ISM}{interstellar medium}
\newacronym{gmc}{GMC}{giant molecular cloud}
\newacronym{kde}{KDE}{kernel density estimator}
\newacronym[firstplural=luminous infrared galaxies (LIRGs)]{lirg}{LIRG}{luminous infrared galaxy}
\newacronym{ms}{MS}{measurement set}
\newacronym{phangs}{PHANGS-ALMA}{Physics at High Angular resolution in Nearby GalaxieS with ALMA}
\newacronym{pdf}{PDF}{probability density function}
\newacronym{ra}{R.A.}{right ascension}
\newacronym{spw}{SPW}{spectral window}
\newacronym{sfe}{SFE}{star formation efficiency}
\newacronym{sfr}{SFR}{star formation rate}
\newacronym{tp}{TP}{total power}
\newacronym{u/lirg}{U/LIRG}{ultra/luminous infrared galaxy}
\newacronym{uv}{UV}{ultra violet}

\title[Molecular cloud properties of the Antennae]{Cloud-Scale Molecular Gas Properties of the Antennae Merger: A Comparative Study with PHANGS-ALMA Galaxies and NGC 3256}

\author[N. Brunetti et al.]{
Nathan Brunetti,$^{1}$ 
Christine D. Wilson,$^{1}$\thanks{E-mail: wilsoncd@mcmaster.ca} 
Hao He,$^{1}$ 
Jiayi Sun,$^{1,2}$ 
Adam K. Leroy,$^{3}$ 
Erik Rosolowsky,$^{4}$ \newauthor
Ashley Bemis,$^{5}$
Frank Bigiel,$^{6}$ 
Brent Groves,$^{7}$ 
Toshiki Saito,$^{8}$
and Eva Schinnerer$^{9}$ 
\\
$^{1}$Department of Physics and Astronomy, McMaster University, Hamilton, ON L8S 4M1, Canada\\
$^{2}$Canadian Institute for Theoretical Astrophysics (CITA), University of Toronto, 60 St George Street, Toronto, ON M5S 3H8, Canada\\
$^{3}$Department of Astronomy, The Ohio State University, 140 West 18th Avenue, Columbus, OH 43210, USA; \\ Center for Cosmology and Astroparticle Physics (CCAPP), 191 West Woodruff Avenue, Columbus, OH 43210, USA\\
$^{4}$Department of Physics, University of Alberta, Edmonton, AB T6G 2E1, Canada\\
$^{5}$Leiden Observatory, Leiden University, PO Box 9513, 2300 RA Leiden, The Netherlands\\
$^{6}$Argelander-Institut für Astronomie, Universität Bonn, Auf dem Hügel 71, D-53121 Bonn, Germany\\
$^{7}$International Centre for Radio Astronomy Research, University of Western Australia, 35 Stirling Highway, Crawley, WA 6009, Australia\\
$^{8}$National Astronomical Observatory of Japan, 2-21-1 Osawa, Mitaka, Tokyo, 181-8588, Japan\\
$^{9}$Max-Planck-Institut für Astronomie, Königstuhl 17, D-69117, Heidelberg, Germany
}

\date{Accepted XXX. Received YYY; in original form ZZZ}

\pubyear{2024}

\begin{document}
\label{firstpage}
\pagerange{\pageref{firstpage}--\pageref{lastpage}}
\maketitle

\begin{abstract}
    We present observations of the central \SI{9}{\kilo\parsec} of the Antennae merger (NGC 4038/9) at \SI{55}{\parsec} resolution in the \acrshort{co} (\num{2}--\num{1}) line obtained with the Atacama Large Millimeter/submillimeter Array (ALMA).
       We use a pixel-based analysis to compare the gas properties in the Antennae to those in 70 nearby spiral galaxies from the \acrshort{phangs} survey, as well as the merger and nearest \acrlong{lirg} \ngc{3256}.
     Compared to \acrshort{phangs} galaxies at matched spatial resolution, the molecular gas in the Antennae exhibits some of the highest surface densities, velocity dispersions, peak brightness temperatures, and turbulent pressures.
    However, the virial parameters in the Antennae are consistent with many of the \acrshort{phangs} galaxies.
   \ngc{3256} has similar gas surface densities but higher nuclear velocity dispersions than the Antennae, as well as higher system-wide peak brightness temperatures and virial parameters.
    \ngc{3256} is at a later stage in the merging process than the Antennae, which may result in more intense merger-driven gas flows that could drive up the turbulence in the gas. The high virial parameters in \ngc{3256} may indicate that this increased turbulence is suppressing future star formation as \ngc{3256} moves out of the starburst phase. In comparison, the relatively normal virial parameters in the Antennae may imply that it is about to undergo a new burst of star formation. 
\end{abstract}




\section{Introduction}
From the spatial correlation of molecular gas and recent star formation seen in galaxies throughout the universe, we know that stars form from molecular gas.
The efficiency with which molecular gas is converted into stars has been observed to be fairly similar at sub-kiloparsec scales across many nearby spiral galaxies \citep{Ler2008}.
However, the global \gls{sfe} (calculated from the \gls{sfr} and molecular gas mass ($\rm M_{\rm mol}$) via $\rm SFE = SFR/M_{\rm mol}$)  
for the most actively star-forming galaxies deviates from what is observed in more normal spiral galaxies \citep[e.g.][]{Ken2021}.
Observations of these different star-forming regimes, done at comparable angular resolutions and sensitivities and using the same tracer, will allow us to understand what causes this difference in star formation efficiency.

Both global and kiloparsec-scale observations of spiral galaxies and starbursts have revealed the difference in \glspl{sfe} but not the cause \citep[e.g.][]{del2019, Wil2019, Ken2021}, so observations of the molecular gas at smaller scales are the natural next step.
Many nearby spiral galaxies have been observed in molecular gas at scales around \SI{100}{\parsec} and even smaller \citep[e.g.][]{Sch2013, Ler2021b}, but the relative rarity of and distance to starbursting galaxies have precluded similar observations.
\citet{Bru2021} and \citet{Bru2022} presented the highest spatial-resolution observations ($\sim$100~pc) of molecular gas in \ngc{3256}, a \gls{lirg} and actively star-forming merger.
Comparisons to the \gls{phangs} homogeneous survey of \num{70} nearby spiral galaxies \citep{Sun2020, Ler2021b} at matched resolution revealed that the molecular gas in the merger reaches some of the highest mass surface densities, velocity dispersions, peak brightness temperatures, virial parameters, and internal turbulent pressures seen in the sample.
However, a single merger that happens to be the nearest \gls{lirg} cannot be expected to be representative of all merging and starbursting systems.
To extend the sample of mergers, we have observed \ngc{} (Arp~\num{244}, ``the Antennae'') at $\sim$100~pc spatial resolution to add it to these comparisons of molecular gas properties.

At a distance of \SI{22}{\mega\parsec} \citep{Sch2008}, \ngc{} is the nearest gas-rich major merger \citep{Sta1990, Wil2000, Gao2001, Wil2003, Bra2009, Ued2012, Sch2014, Big2015} with $\sim$\SI{2e10}{\solarmass} of molecular gas in the central region. 
Based on a significant body of numerical work on reproducing the morphology and kinematics of the interaction, the system is currently either just before or just after the second pericentre passage \citep[e.g.][]{Too1972, Bar1988, Mih1993, Kar2010, Tey2010, Pri2013, Ren2015}.
Its central region hosts two progenitor nuclei separated by about \SI{7}{\kilo\parsec}.
Given these orbital details, \ngc{} is likely at an earlier merger stage than \ngc{3256}, and so offers a chance to probe how the molecular gas properties depend on the phase of the merger.

The total \gls{sfr} of the Antennae is between \SI{11}{\solarmass\per\year} (separately estimated from far-ultraviolet probing $\sim$\SIrange{1}{100}{\mega\year} and \SI{24}{\micro\metre} probing $\sim$\SIrange{1}{400}{\mega\year}) and \SI{20}{\solarmass\per\year} \citep[from H$\alpha$ probing $\sim$\SIrange{1}{10}{\mega\year};][]{Cha2017}. Although not technically a \gls{lirg} from its infrared luminosity, the higher of these \gls{sfr} estimates would give the Antennae a similar \gls{sfr} to LIRGs.
Separated from the nuclei is the starbursting ``overlap region'' which alone exhibits a \gls{sfr} of about \SI{4}{\solarmass\per\year} over the last $\lesssim$\SI[input-comparators=\lesssim]{100}{\mega\year} estimated from mid- through far-\acrlong{ir} observations \citep{Bra2009, Kla2010, Bem2019}. The remainder of the central part of the merger has an infrared-estimated \gls{sfr} totaling about \SI{2.6}{\solarmass\per\year} \citep{Bem2019}.

\ngc{} hosts an estimated population of at least \num{1e4} young massive star clusters, with masses in some cases exceeding \SI{e6}{\solarmass} \citep{Whi2010, Cha2015, Mok2020}.
Several stellar populations of different ages have also been identified throughout the system with a young starburst population in the overlap region \citep[$\sim$\SIrange{3}{10}{\mega\year};][]{Men2001, Men2005, Whi2010} and older post-starburst populations in the nuclei \citep[$\sim$\SI{65}{\mega\year};][]{Men2001}.
Exploring the molecular-gas properties of this system will provide the details on what conditions are necessary to form the most massive star clusters \citep[e.g.][]{He2022}.
Our understanding of nearby mergers like \ngc{} should also fill in the small-scale information on how star and cluster formation occurred at high redshift, where mergers were much more common \citep[e.g.][]{Rom2021} and when the progenitors of globular clusters formed.

In this paper we present \gls{alma} observations of the central $\sim$\SI{9}{\kilo\parsec} of \ngc{} in the \gls{co} $J$=\num{2}--\num{1} line at linear scales of 55 pc, corresponding to \gls{gmc}-scales, and use them
to probe the molecular gas properties across the diverse \gls{ism}  of \ngc{}. These observations have comparable sensitivity and resolution but are a significant improvement in area coverage (see Figure~\ref{fig:moment_maps_55}), and image fidelity (covering the full range of spatial scales) over earlier \gls{alma} observations of the Antennae \citep{Whi2014,Sun2018}. 
Section~\ref{sec:data} describes the new observations, along with our calibration and imaging procedure for the \gls{co} (\num{2}--\num{1}) data.
In Section~\ref{sec:analysis} we describe the pixel-based method used to measure the molecular gas properties in \ngc{} at a range of spatial resolutions.
In Section~\ref{sec:results}, we compare \ngc{} to the \gls{phangs} results presented by \citet{Sun2018, Sun2020} and the \gls{lirg} NGC 3256 presented in \citet{Bru2021}, and we discuss the implications in Section~\ref{sec:discussion}.
Finally, in Section~\ref{sec:conclusions} we summarize the results and conclusions of this work.

\section{Data} \label{sec:data}
\subsection{Observations}
\begin{figure*}
    \centering
     \includegraphics[scale=0.8]{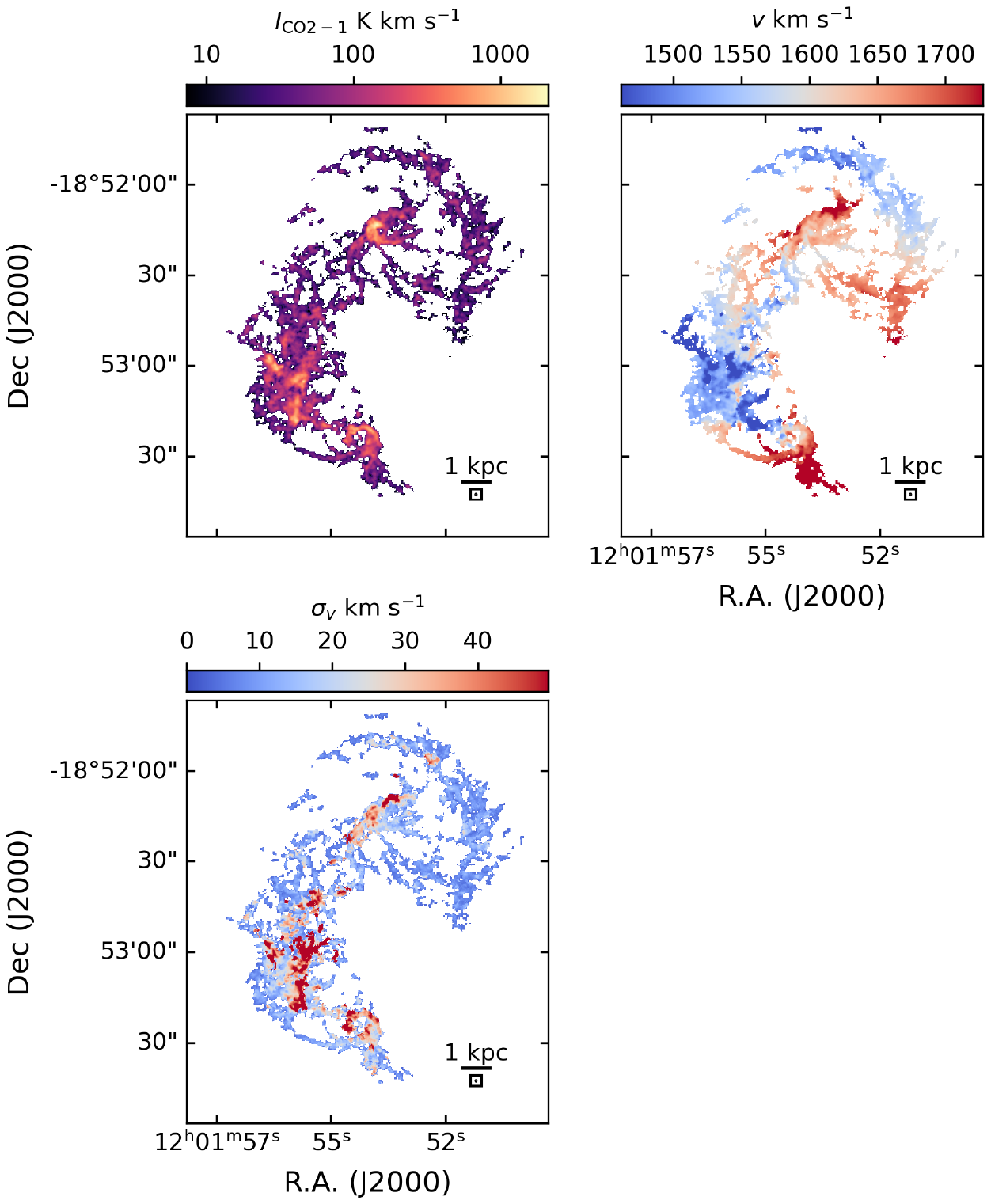}
   \caption{
        \gls{co} (\num{2}--\num{1}) moment maps of \protect\ngc{} at \SI{55}{\parsec} resolution. Top left: integrated intensity (moment 0); top right: velocity field (moment 1); bottom: velocity dispersion (moment 2).
        Pixels are half of the beam \gls{fwhm} on a side.
                The squares in the bottom right corner of each panel contain a circle with diameter equal to the beam \gls{fwhm} (0.51$^{\prime\prime}$). A scale bar indicating \SI{1}{\kilo\parsec} at the distance of \protect\ngc{} is also shown in the bottom right corner of each panel. 
    }
    \label{fig:moment_maps_55}
\end{figure*}

Mosaicked \gls{alma} observations of the central $\sim$\SI{9}{\kilo\parsec} of \ngc{} covering the entire disks of the two merging galaxies were obtained in Cycle \num{6} (see Table~\ref{tab:observations} for details of the observations).
The main \SI{12}{\metre}-array was used in both a compact and extended configuration.
The \gls{aca} was included for sensitivity to larger-scale emission, and the \gls{tp} array was also included to capture the largest-scale emission.
Band \num{3} and \num{6} observations were carried out with spectral setups designed to cover the $J$=\num{1}--\num{0} and $J$=\num{2}--\num{1} rotational transitions of $^{12}$\gls{co}, the $J$=\num{1}--\num{0} transitions of C$^{17}$O and \gls{cn}, and the $J$=\num{2}--\num{1} transitions of $^{13}$\gls{co} and C$^{18}$O.
Sufficient bandwidth was covered to also allow detection of continuum emission in both bands.
The spectral resolution of the $^{12}$\gls{co} (\num{2}--\num{1}) spectral window is \SI{1.953}{\mega\hertz}, or approximately \SI{2.54}{\kilo\metre\per\second}. The native beam size is 0.51$^{\prime\prime}$ (55 pc) and the rms noise at this resolution is 0.24 K. The calibration accuracy of ALMA at these frequencies is 5\%.This paper focuses on the $^{12}$\gls{co} $J$=\num{2}--\num{1} line; the continuum data products have been presented in \citet{He2022}. The remaining CO spectral lines are presented in He et al. 2024 (submitted).

\subsection{Calibration and imaging}
Calibration of all interferometric data was carried out by observatory staff as part of data quality assurance using the \gls{alma} pipeline for \gls{casa} \citep{McM2007};
\gls{casa} and pipeline versions used for calibration are summarized in Table~\ref{tab:observations}. After downloading the raw data from the archive, we used the same versions to reapply the calibration prior to our imaging.
We inspected the diagnostic plots in the observatory web-logs to search for problematic data that were not properly calibrated or left unflagged by the \gls{alma} pipeline.
Nothing was identified that was serious enough to warrant changes to the calibration or additional flagging.

Given the complex morphology of molecular line emission expected in the Antennae at the resolution of our observations, and the number of lines to be imaged, we requested access to an early version of the \gls{phangs} imaging pipeline\footnote{The procedure followed by the imaging pipeline has not changed dramatically between version one and the version described in detail by \citet{Ler2021a}.} for \gls{casa} \citep{Ler2021a}.
Starting from the version-one code (at commit 5ef53d3), we made modifications for imaging the additional array combinations and spectral lines included in our observations of \ngc{}\footnote{
    Details of these modifications are given in Appendix~\ref{sec:pipeline_modifications_appendix}.
    Our modified version of the imaging pipeline will be shared on reasonable request to the corresponding author, with a fully annotated \textsc{Git} change history.
}.
These modifications include implementation of different angular sizes of the model components during multi-scale cleaning, for which we use point sources, 0.5, 1, 2.5, 5, and 10 arcsec.
All spectral-line cleaning was carried out using the newest \gls{casa} version available at the time (5.6.1-8) to run the \gls{phangs} imaging pipeline. The observations from the two \SI{12}{\metre}-array configurations and the \gls{aca} were all imaged and cleaned together.
Cleaned cubes were also visually inspected for artefacts related to calibration and cleaning errors before combining with the \gls{tp} data.

\Gls{tp} data processing was carried out with the \gls{phangs} \gls{tp} calibration and imaging pipeline for \gls{casa} in \gls{casa} version 4.7.2 (see Table~\ref{tab:observations} for more details).
This version of \gls{casa} was the newest version that the \gls{tp} pipeline had been tested with at the time.
\cite{Her2020} describe the full details of the pipeline procedure.
We used a modified version that does not perform any spectral binning and
all continuum subtraction was carried out using order-one polynomial (linear) fits to the continuum levels.

The interferometric and \gls{tp} cubes were combined using the \gls{phangs} imaging pipeline to produce the complete measurements of the spectral line emission in \ngc{} (again using \gls{casa} version 5.6.1-8).
Interferometric cubes corrected for the primary beam response were first made by dividing by the primary beam cubes, and the corrected cubes were convolved to have circular synthesized beams.
The interferometric cubes were padded with masked pixels to cover at least the entire \gls{tp} \gls{fov}, and the \gls{tp} cubes were regridded on to the same astrometric and spectral grid as the interferometric cubes.
Using the \emph{feather} task in \gls{casa}, the \gls{tp} cubes were combined with both main array configurations plus the \gls{aca}.
The feathered cubes still contain the primary beam response applied to the interferometric data (and the inherent response in the \gls{tp} data), so another copy of the cubes was made which was multiplied by the interferometric primary beam response to produce flat-noise cubes for future use in identifying individual clouds using algorithms such as \textsc{pycprops} \citep{Ros2021}.

Several versions of the interferometric cube were made when the synthesized beam was made circular, with \gls{fwhm} physical sizes of \SIlist{55; 80; 90; 120; 150}{\parsec}.
These beam sizes facilitate direct comparisons to the analyses at \SIlist{55; 80; 120}{\parsec} by \citet{Sun2018} and \citet{Bru2021}, as well as at \SIlist{90; 150}{\parsec} by \citet{Sun2020} and \citet{Bru2022}. The cube  with \SI{55}{\parsec} resolution (\SI{0.51}{\arcsec}) has an rms noise of approximately 0.25 K.
Following \citet{Sun2018} and \citet{Bru2021}, after feathering the \gls{tp} data with the convolved interferometric cubes, each cube was spatially regridded such that the pixels were half the synthesized beam \gls{fwhm} on a side to roughly Nyquist sample the beams. 
Moment maps calculated from the  \gls{co} (\num{2}--\num{1}) data cube as described in \citet{Bru2021} are shown in 
Figure~\ref{fig:moment_maps_55}.

\section{Analysis} \label{sec:analysis}
\subsection{Measuring molecular gas properties} \label{sec:measuring_properties}

Maps of total molecular gas mass surface density ($\Sigma_{\mathrm{mol}}$), velocity dispersion ($\sigma_{v}$), and peak brightness temperature ($T_{\mathrm{B, max}}$) were calculated in the same way as in \citet{Bru2021}.
The only difference was that we adopted a single Milky-Way like conversion factor of $\alpha_{\mathrm{CO(2-1)}} = 4.35 / 0.70 = \SI{6.25}{\solarmass\per\square\parsec}(\si{\kelvin\kilo\metre\per\second})^{-1}$ for all measurements from \ngc{}, following \citet{Sun2018}, where 0.70 is the commonly used value for the $^{12}$\gls{co} $J$=\num{2}--\num{1} / $J$=\num{1}--\num{0}  line ratio \citep[cf.][]{Leroy2022}{}{}.
Using the same conversion factor as \citet{Sun2018} means any differences in the estimated mass surface densities originate from different integrated intensities.
The conversion factor in \ngc{} is not yet well constrained (see Section~\ref{sec:discussion} for further discussion) but previous investigation has estimated it may be similar to the typical Milky Way value \citep{Wil2003}.
This conversion factor includes a contribution from helium and other heavy elements.

\begin{figure*}
    \centering
    \includegraphics{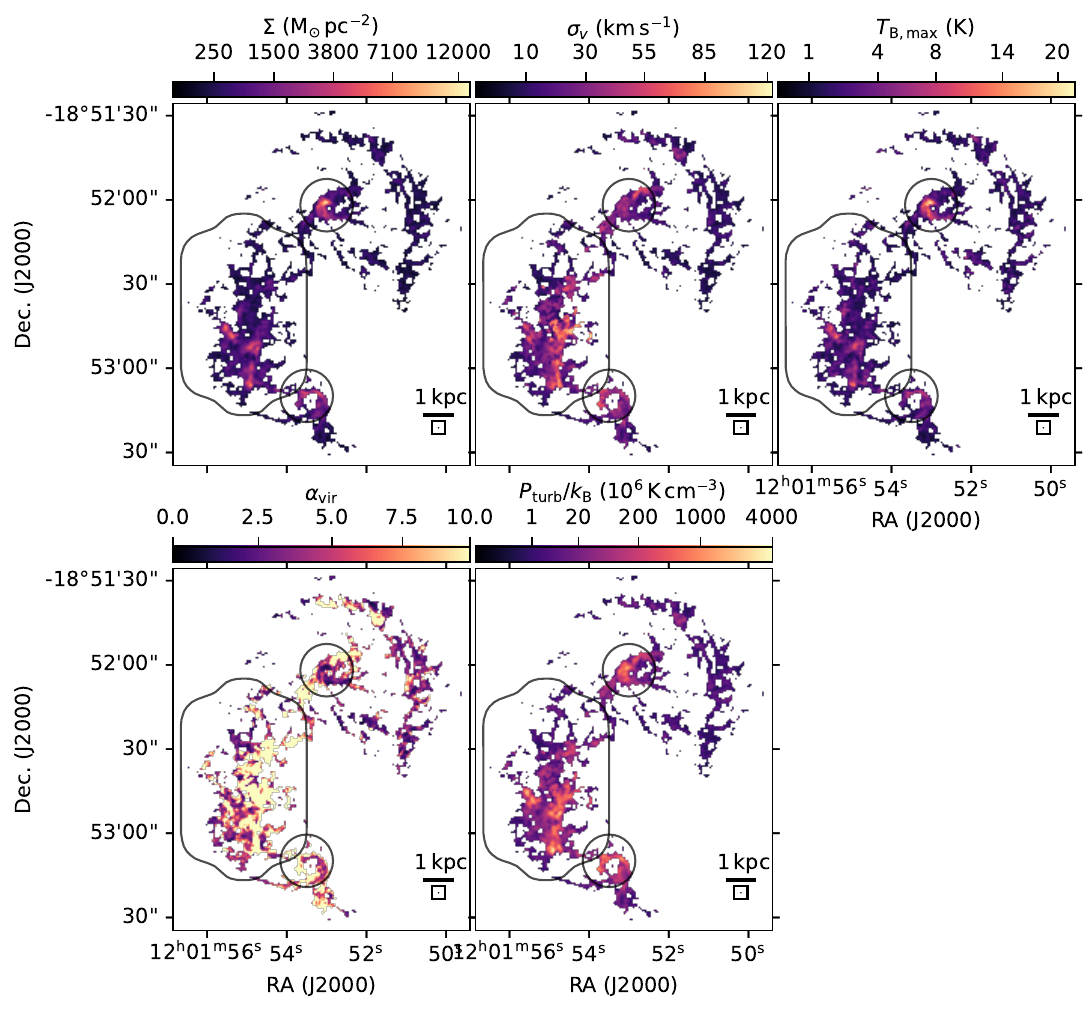}
   \caption{
        Maps of molecular-gas properties in \protect\ngc{} at \SI{55}{\parsec} resolution, calculated from \gls{co} (\num{2}--\num{1}) moment maps.
        The top row shows, from left to right, mass surface density, velocity dispersion, and peak brightness temperature.
        The bottom row shows the virial parameter on the left and the internal turbulent pressure on the right.
        Pixels are half of the beam \gls{fwhm} on a side.
        Given the range of measured virial parameters, we chose a relatively low upper clipping value of $\alpha_{\mathrm{vir}} = 10$ to show pixels that are likely unbound (yellow) while still showing some of the structure for pixels that may be bound by external material (orange) and pixels that are likely marginally bound (black/blue).
        The gray circles indicate the regions within \SI{1}{\kilo\parsec} radii of the nuclei, denoting the nuclear and non-nuclear regions discussed in later sections.
        The gray polygons in the lower left indicate the overlap-region \gls{fov} observed in \gls{co} (\num{3}--\num{2}) by \citet{Whi2014} and analysed by \citet{Sun2018}.
        Pixels in these polygons make up our overlap region sub-sample.
        The squares in the bottom right corner of each panel contain a circle with diameter equal to the beam \gls{fwhm} (0.51$^{\prime\prime}$).
        A scale bar indicating \SI{1}{\kilo\parsec} at the distance of \protect\ngc{} is also shown in the bottom right corner of each panel.
        A \gls{co}-to-H$_{2}$ conversion factor of \SI{6.25}{\solarmass\per\square\parsec}(\si{\kelvin\kilo\metre\per\second})$^{-1}$ was used to convert integrated intensities to mass surface densities, which also affects the estimates of the virial parameter and internal turbulent pressure (see Equations~\ref{eq:alpha_vir} and \ref{eq:p_turb}). 
    }
    \label{fig:maps_55}
\end{figure*}

From our measurements of $\Sigma_{\mathrm{mol}}$ and $\sigma_{v}$ we also estimate the virial parameter and internal turbulent pressure in each pixel, following \citet{Sun2020}, with
\begin{equation} \label{eq:alpha_vir}
    \alpha_{\mathrm{vir}} \approx 3.1 \left(\frac{\Sigma_{\mathrm{mol}}}{\SI{1e2}{\solarmass\per\square\parsec}}\right)^{-1} \left(\frac{\sigma_{v}}{\SI{10}{\kilo\metre\per\second}}\right)^{2} \left(\frac{D_{\mathrm{beam}}}{\SI{150}{\parsec}}\right)^{-1}
\end{equation}
and
\begin{equation} \label{eq:p_turb}
    \frac{P_{\mathrm{turb}}}{k_{\mathrm{B}}} \approx \SI{3.3e5}{\kelvin\per\cubic\centi\metre} \left(\frac{\Sigma_{\mathrm{mol}}}{\SI{1e2}{\solarmass\per\square\parsec}}\right) \left(\frac{\sigma_{v}}{\SI{10}{\kilo\metre\per\second}}\right)^{2} \left(\frac{D_{\mathrm{beam}}}{\SI{150}{\parsec}}\right)^{-1}.
\end{equation}
where $D_{\mathrm{beam}}$ is the \gls{fwhm} of the synthesized beam.
These are equivalent to equations~\num{13} and \num{15} from \citet{Sun2018} that were used by \citet{Bru2021}.
As described in \citet{Sun2018}, these equations assume that a single cloud roughly fills each synthesized beam along with the equation for the virial parameter assuming a cloud density profile that goes as $\rho(r) \propto r^{-1}$.
Maps of molecular gas properties at \SIlist{55}{\parsec} resolution are shown in Figure~\ref{fig:maps_55}; maps at \SIlist{150}{\parsec} resolution are shown in
Figure~\ref{fig:maps_150}.
We provide the binned data used in this paper for both \ngc{} and \ngc{3256} as two separate machine-readable tables (see Table~\ref{table:antennae_pixel_data} and~\ref{table:ngc3256_pixel_data} for format). Note that we use square pixels, following \citet{Sun2018} and \citet{Bru2021}, but \citet{Sun2020} resample their maps to have hexagonal pixels that match the beam size before measuring the gas properties. 

\subsection{Separating measurements by region}
To examine the impact that the location within \ngc{} has on the molecular gas properties, we separate our measurements into sub-samples that are close to the nuclei and farther away.
Following \citet{Sun2018} and \citet{Bru2021}, all pixels that are less than \SI{1}{\kilo\parsec} from either nucleus (shown as gray circles with radii of \SI{1}{\kilo\parsec} in Figures~\ref{fig:maps_55} and \ref{fig:maps_150}) are included in the nuclear pixel sample, and the remaining pixels make up the non-nuclear sample.
Since the \gls{co} (\num{3}--\num{2}) observations of \ngc{} from \citet{Whi2014} analysed by \citet{Sun2018} only covered the overlap region, we also separate pixels in our \gls{co} (\num{2}--\num{1}) data into another sub-sample that lies in the same on-sky footprint  as the \gls{co} (\num{3}--\num{2}) observations for direct comparisons between the two datasets (gray polygons in Figures~\ref{fig:maps_55} and \ref{fig:maps_150}).
Note that \citet{Sun2020} separate their aperture measurements
into central and disc regions within each galaxy based on distinct structures identified in near-infrared images \citep[the full description of this methodology is given by][]{Que2021} rather than by a 1 kpc radius as we do in this work\footnote{
    Table of measurements from \citet{Sun2020} was retrieved from the journal website on 2021 September 6.
}.

\subsection{Morphology of the CO emission at 55 pc resolution}

The CO emission of the Antennae shown in Figure~\ref{fig:moment_maps_55} has a very complex structure. As seen in earlier work \citep{Sta1990,Wil2000,Ued2012}, the regions of highest integrated intensity in the moment 0 map are found in the northern nucleus (NGC~4038), the southern nucleus (NGC~4039), and the overlap region. These regions with high integrated intensities also typically show large line widths in the moment 2 map and high brightness temperatures and pressures (Figure~\ref{fig:maps_55}). Both nuclei show rather distorted morphologies, especially NGC~4039, with filaments at different pitch angles emerging from each nucleus.

The velocity field shown in the moment 1 map is particularly striking in appearance, with sudden shifts in velocity on the scale of a few resolutions elements visible in many parts of the map. The western arm (to the west of NGC~4038) shows a consistent velocity gradient reminiscent of the velocity field of a rotating disk, but this interpretation is hard to reconcile with the quite redshifted emission from the nucleus of NGC~4038 itself. The emission in the central 40 arcsec of NGC~4039 also suggests large-scale rotation, but this organized structure breaks up in the central 5-10 arcsec, where several shifts from red- to blue-shifted emission are seen.

There are a variety of filamentary structures visible in the integrated intensity map, including the long filament to the east of NGC~4039 highlighted in \citet{Espada2012}, and also a variety of other similar structures particularly to the south of the NGC~4038 nucleus. While these structures typically exhibit narrow line widths in the moment 2 map, some of them show a nearly constant radial velocity in the moment 1 map while others shows signs of a gradient in radial velocity along their length.

\section{Results} \label{sec:results}
In this section, we compare the molecular gas properties of \ngc{} with spiral galaxies from the \gls{phangs} survey as well as with the merger and nearest \gls{lirg} \ngc{3256}. In Section~\ref{sec:samples} we present the mean values for gas mass surface density, velocity dispersion, and peak brightness temperature along with two derived quantities, the virial parameter and the turbulent pressure. In this analysis, we distinguish between the central nuclear regions and the rest of the disk. In Section~\ref{subsec-properties}, we examine the correlations between velocity dispersion and mass surface density in the various galaxies and regions.

\subsection{Distribution of cloud-scale molecular gas properties} \label{sec:samples}
\begin{figure}
    \centering
    \includegraphics{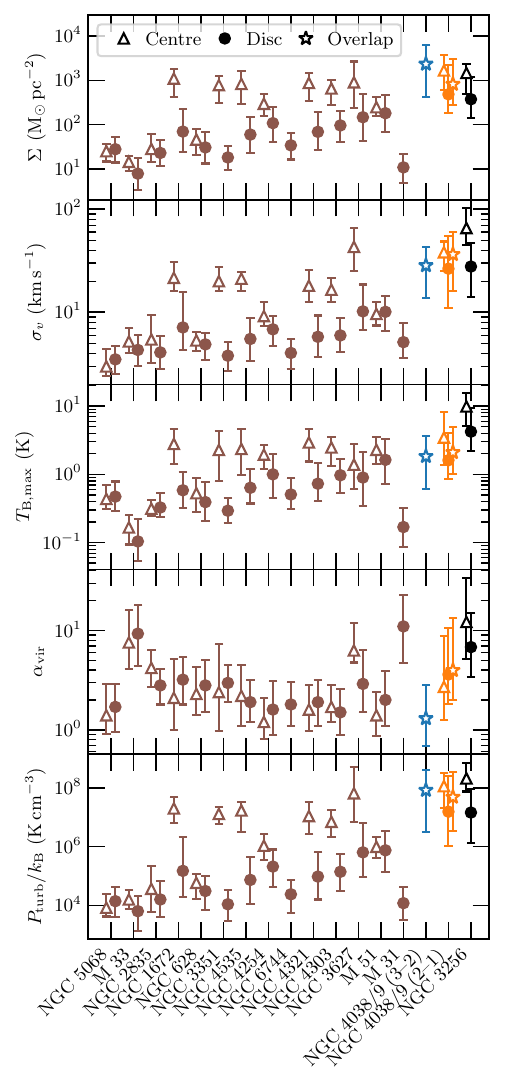}
    \caption{
        Comparisons of mass-weighted pixel sample percentiles at \SI{120}{\parsec} resolution from \protect\ngc{}, \protect\ngc{3256}, and \citet{Sun2018} for mass surface density, velocity dispersion, peak brightness temperature, virial parameter, and internal turbulent pressure.
        Medians are shown as symbols and the inner \nth{68} percentiles as error bars.
        Samples are split between centre (triangle) and non-centre/disc (circle) pixels.
        \protect\ngc{} also has pixel samples from the \citet{Whi2014} overlap region \gls{fov}, shown as stars.
        Percentiles from our \gls{co} (\num{2}--\num{1}) observations of \protect\ngc{} are in orange, from the \gls{co} (\num{3}--\num{2}) observations in blue, \protect\ngc{3256} is in black, and the remainder of the galaxies analysed by \citet{Sun2018} are in brown.
        Galaxies from \citet{Sun2018} are ordered along the $x$ axis with stellar mass increasing from left to right.
        Stellar masses are from \citet{Sun2020}, where available, and from \citet{Sun2018} otherwise (i.e. M~\num{31}, M~\num{33}, and M~\num{51}).
    }
    \label{fig:percentiles_sun_2018}
\end{figure}

\begin{figure}
    \centering
    \includegraphics{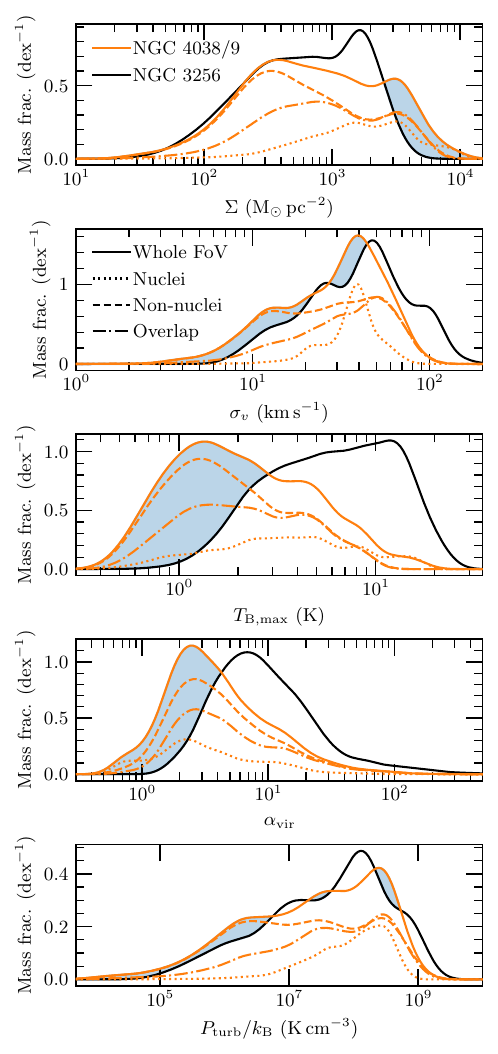}
    \caption{
     Mass-weighted Gaussian \glspl{kde} of the \SI{120}{\parsec} resolution measurements from the entire \glspl{fov} of \protect\ngc{} and \protect\ngc{3256} in the solid orange and black lines, respectively.
        The samples from \protect\ngc{} are also separated into regions near the nuclei (dotted), farther than \SI{1}{\kilo\parsec} from the nuclei (dashed), and within the \citet{Whi2014} overlap region \gls{fov} (dash-dotted).
        The blue-shaded regions show where the whole-\gls{fov} sample from \protect\ngc{} is above the sample from \protect\ngc{3256}.
        Gaussian \gls{kde} bandwidths for the whole \gls{fov} samples were automatically chosen using the \textsc{scipy} implementation of Scott's Rule \citep{Sco1992}, and the other regions from \protect\ngc{} use the same bandwidth as that of the whole \gls{fov}. 
    }
    \label{fig:1d_kdes_mergers}
\end{figure}

Qualitative comparisons between \ngc{}, \ngc{3256}, and the \gls{phangs} galaxies presented by \citet{Sun2018} at \SI{120}{\parsec} resolution are shown as mass-weighted medians (symbols) and inner \nth{68} percentiles (errorbars) in Figure~\ref{fig:percentiles_sun_2018}.
Galaxies are split between nuclear (triangles) and non-nuclear (circles) sub-samples, with the \citet{Whi2014} overlap region \gls{fov} also included for \ngc{} (stars).
Figure~\ref{fig:1d_kdes_mergers} shows the mass-weighted distributions of the various sub-samples from \ngc{} and \ngc{3256} smoothed with Gaussian kernel density estimators 
\citep[KDEs, e.g.][]{Sun2018}{}{}\footnote{Measurements at \SI{120}{\parsec} resolution are shown but comparisons at \SI{80}{\parsec} and \SI{55}{\parsec} resolution were also made with generally the same results as seen in Figures~\ref{fig:percentiles_sun_2018} and \ref{fig:1d_kdes_mergers}.}.

There is considerable overlap in all cloud-scale properties between the nuclear and non-nuclear sub-samples from \ngc{}. \ngc{} exhibits higher mass surface densities measured in \gls{co} (\num{2}--\num{1}) relative to  all the \gls{phangs} galaxies\footnote{We note that the surface density values derived from the \gls{co} (\num{3}--\num{2}) measurements for \ngc{} from \citet{Sun2018} exceed the \gls{co} (\num{2}--\num{1}) values. This difference is likely due to the assumed value for the \gls{co} $J$=\num{3}--\num{2} / $J$=\num{1}--\num{0} line ratio being smaller than the value observed in He et al. 2024 (submitted).} in Figure~\ref{fig:percentiles_sun_2018}.
The nuclear and non-nuclear surface density samples from \ngc{} and \ngc{3256} are similar.
Mass surface density distributions from the central regions of \gls{phangs} galaxies overlap the most with \ngc{}, but the disc region distributions from several galaxies also overlap with \ngc{}.
However when comparing nuclei to nuclei and non-nuclei to non-nuclei, the samples from \ngc{} are wider, have slightly higher medians, and extend to higher surface densities at all resolutions.
The greater sample spread and maximum surface densities stand out especially in Figure~\ref{fig:1d_kdes_mergers}.

The results of comparing \ngc{} to the \gls{phangs} galaxies in velocity dispersion, peak brightness temperature, and internal turbulent pressure are similar to those seen in mass surface density, with 
 the central regions of the PHANGS-ALMA galaxies overlapping the most with NGC 4038/9.
\ngc{} exhibits some of the highest values, although the centres of \ngc{3256} and sometimes \ngc{3627} often reach significantly higher values.
\ngc{} appears most consistent with the \gls{phangs} galaxies in virial parameter, with the relatively low median and very large sample width meaning it overlaps at least partially with all \gls{phangs} galaxies. 
The higher mass surface densities and slightly lower velocity dispersions measured in \gls{co} (\num{3}--\num{2}) in \ngc{} lead to virial parameter samples from the \gls{co} (\num{3}--\num{2}) measurements that are much lower than in \gls{co} (\num{2}--\num{1}) via Equation~\ref{eq:alpha_vir}.
However, Equation~\ref{eq:p_turb} results in the \gls{co} (\num{3}--\num{2}) sample of internal turbulent pressures being only slightly higher than in \gls{co} (\num{2}--\num{1}).
Peak brightness temperatures are also similar between the two spectral line samples.

In properties other than mass surface density, the medians from \ngc{} are often lower than those measured in \ngc{3256}.
The two mergers appear most consistent in their non-nuclear samples of velocity dispersion and pressure, though the sample from \ngc{} is wider, extending both above and below the sample from \ngc{3256}.
The comparison of virial parameters in \ngc{} and \ngc{3256} is quite different from the other gas properties.
The median virial parameter in \ngc{} is less than half that in \ngc{3256}.
The nuclear median from \ngc{} also appears lower than that of the non-nuclear pixels, contrary to \ngc{3256} but similar to what is seen in the centres and discs of the \gls{phangs} galaxies.
However, the spread of the samples from both \ngc{} and \ngc{3256} does result in considerable overlap in the virial parameter between the two mergers.

\begin{figure*}
    \centering
    \includegraphics{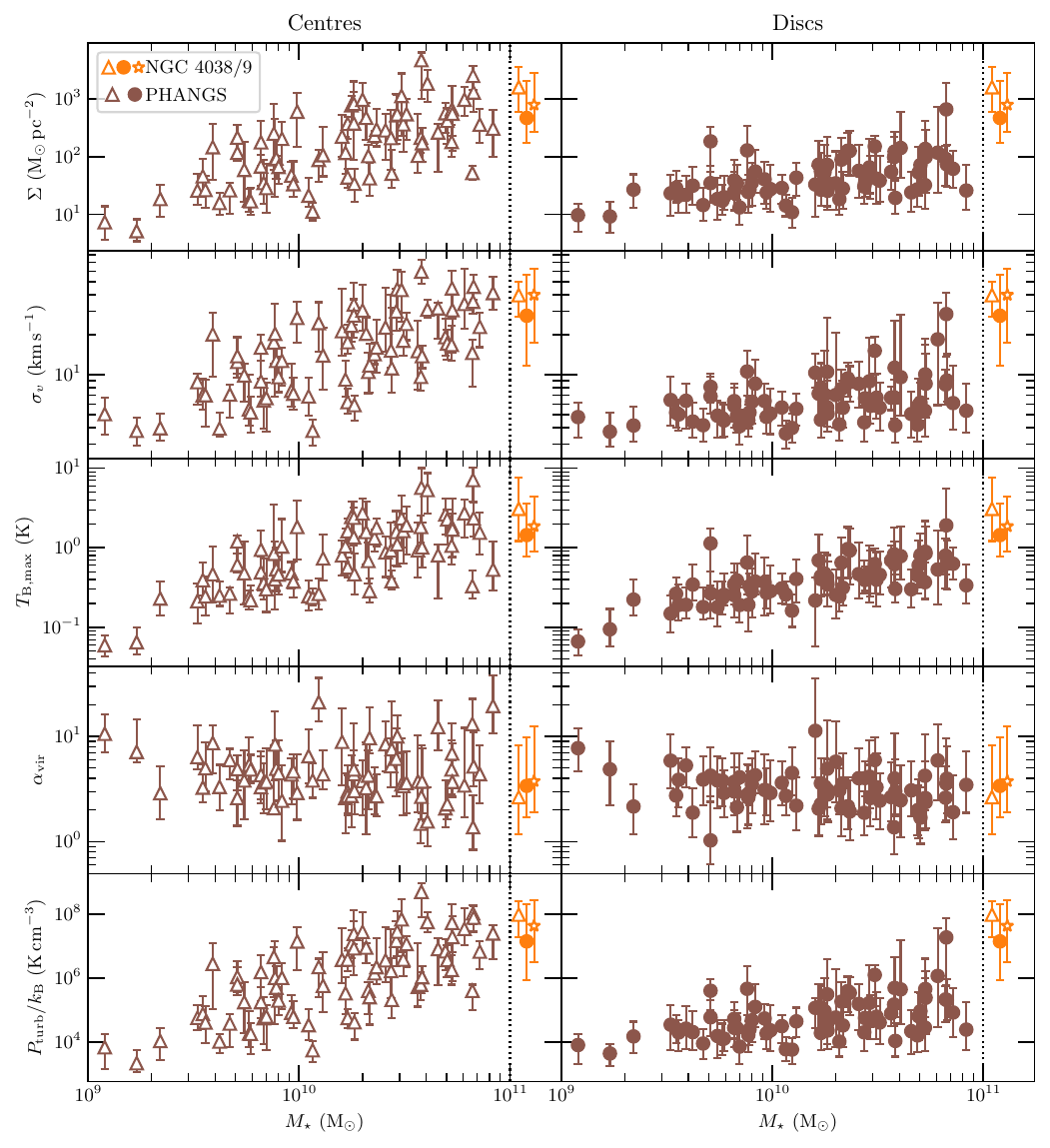}
    \caption{
        Mass-weighted pixel-sample percentiles vs. stellar mass for the five quantities from Fig.~\ref{fig:percentiles_sun_2018} from \protect\ngc{} (orange) and \citet{Sun2020} (brown) measured at \SI{150}{\parsec} resolution.
        The left column shows \gls{phangs} samples from galaxy centres (triangles) and the right column shows samples from outside the centres (circles; based on being within a \SI{1}{\kilo\parsec} radius of the centre, not the centre and disc designations from \citealt{Sun2020}).
        \protect\ngc{} samples from the nuclei, outside the nuclei (including the overlap region), and just the overlap region (stars) are shown in both columns.
        Note that the position of \protect\ngc{} along the $x$ axis is arbitrary.
    }
    \label{fig:percentiles_sun_2020}
\end{figure*}

The same comparisons of physical properties are made in Figure~\ref{fig:percentiles_sun_2020} but the samples from \ngc{} are now shown with the full \gls{phangs} sample  analysed by \citet{Sun2020} at \SI{150}{\parsec} resolution\footnote{We found very little difference between using measurements made at \SI{90}{\parsec} and \SI{150}{\parsec} resolution.}.
Note that we made the separation of pixels into regions in the \gls{phangs} galaxies in the same manner as for \ngc{}, not using the centre and disc designations from \citet{Sun2020}.
We chose this approach to reduce variability in the comparison with \ngc{} potentially brought on by differences in how pixels are assigned to the different regions.
Using the simple \SI{1}{\kilo\parsec} radius also has the benefit of often increasing the number of pixels within the centres of galaxies, resulting in less stochastic inner \nth{68} percentiles for the less massive and/or more distant galaxies.
However, given the size-stellar mass relation of galaxies, the fraction of a galaxy included in the central region will increase as stellar mass decreases.
For example, across the stellar mass range  (\SIrange{1.2e9}{8.3e10}{\solarmass}) of the \gls{phangs} galaxies shown in Figure~\ref{fig:percentiles_sun_2020}, a range of stellar radii of about \SIrange{6}{30}{\kilo\parsec} is expected \citep{Tru2020, San2020}.
This effect will likely result in the percentiles from the centres of low stellar mass galaxies being underestimated since contaminating disc pixels will make up a larger fraction of pixels in their central \SI{1}{\kilo\parsec}.

Trends in the medians and inner \nth{68} percentiles with resolution are not very strong in \ngc{} so the GMC measurements for the Antennae in Figure~\ref{fig:percentiles_sun_2020} do not appear very different compared to Figure~\ref{fig:percentiles_sun_2018}.
The general trends between \ngc{} and \gls{phangs} disc galaxies observed in Figure~\ref{fig:percentiles_sun_2018} appear in the comparisons  in Figure~\ref{fig:percentiles_sun_2020} as well.
The centres of \gls{phangs} galaxies are the most similar to \ngc{}, regardless of spatial scale, with greater similarities
at higher stellar masses.
\ngc{} exhibits some of the highest mass surface densities, velocity dispersions, peak brightness temperatures, and internal turbulent pressures compared to the \gls{phangs} galaxies.
Virial parameters measured in \ngc{} are much more similar to typical values measured in \gls{phangs} galaxies.
Finally, for each property shown in Figure~\ref{fig:percentiles_sun_2020}, there are usually at least a few \gls{phangs} galaxies that are more extreme in one or more aspects than \ngc{}. 

\subsection{Correlations between velocity dispersion and mass surface density}
\label{subsec-properties}

\begin{figure}
    \centering
    \includegraphics{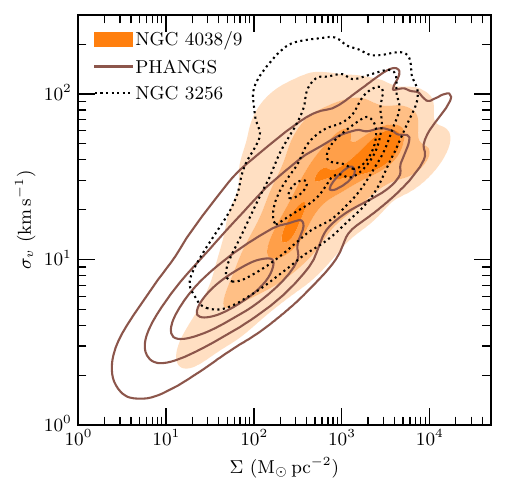}
    \caption{
        Mass-weighted Gaussian \glspl{kde} of velocity dispersion vs. mass surface density measured in all significant pixels or hexagonal apertures from \protect\ngc{} (orange filled contours), \protect\ngc{3256} (black dotted contours), and all \gls{phangs} galaxies presented by \citet{Sun2020} (brown contours).
        Contours enclose \numlist{99.5; 90; 50; 20} per cent of the mass from each data set.
        Measurements shown are made at \SI{150}{\parsec} resolution in \protect\ngc{} and \gls{phangs} and \SI{120}{\parsec} resolution in \protect\ngc{3256}. Note that the mass weighting used in this figure results in a different distribution of contours compared to those presented in \citet{Sun2018,Sun2020}, which did not use mass weighting. 
    }
    \label{fig:dispersion_vs_surface_density_whole}
\end{figure}

\begin{figure}
    \centering
    \includegraphics{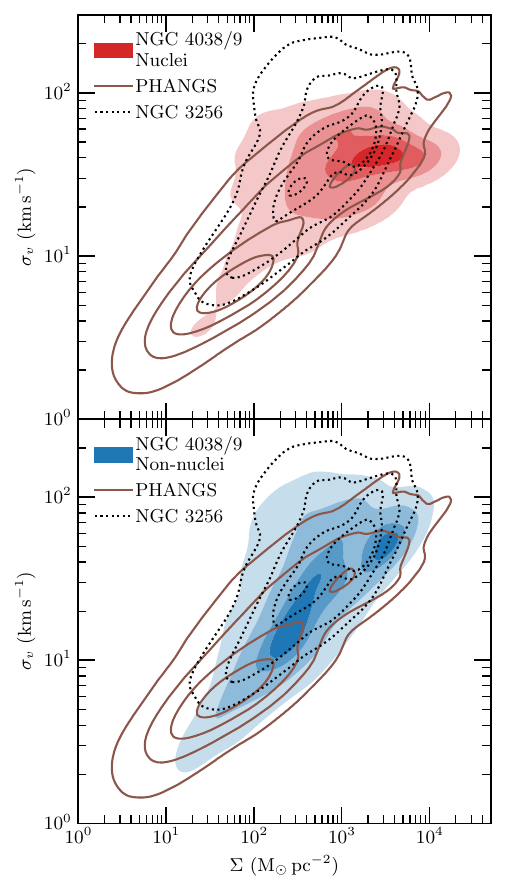}
    \caption{
        Same as Figure~\ref{fig:dispersion_vs_surface_density_whole} but now the sample from \protect\ngc{} is separated into pixels within \SI{1}{\kilo\parsec} of the nuclei (red) and those beyond (blue).
        Contours from \protect\ngc{3256} and \gls{phangs} are the same as in Figure~\ref{fig:dispersion_vs_surface_density_whole} and are identical in the top and bottom panels of this figure.
    }
    \label{fig:dispersion_vs_surface_density_split}
\end{figure}

Figures~\ref{fig:dispersion_vs_surface_density_whole} 
and ~\ref{fig:dispersion_vs_surface_density_split}
show mass-weighted \glspl{pdf} of the velocity dispersion versus mass surface density.
All \glspl{pdf} are made from Gaussian \glspl{kde}, with bandwidths automatically chosen using the \textsc{scipy} implementation of Scott's Rule \citep{Sco1992}.

Figure~\ref{fig:dispersion_vs_surface_density_whole} shows velocity dispersion vs. mass surface density \glspl{pdf} of all significant pixels from \ngc{}, \ngc{3256}, and apertures from all galaxies presented by \citet{Sun2020}.
\glspl{kde} show measurements made at \SI{150}{\parsec} resolution in \ngc{} and \gls{phangs} and \SI{120}{\parsec} resolution in \ngc{3256}\footnote{Again, there are very few qualitative differences between different resolution versions of these figures.}.
The shift of most of the mass to both higher surface densities and velocity dispersions in \ngc{} and \ngc{3256} is visible, but the lower values from the merger samples overlap with the higher values from the \gls{phangs} galaxies.
The apertures from the centres of (primarily barred) \gls{phangs} galaxies populate a very similar part of the parameter space to \ngc{} and \ngc{3256} (see Figure~\num{2} from \citealt{Sun2020} for the separation of \gls{phangs} samples by region and whether a galaxy contains a bar).
These central values make a dispersion-surface density trend that is slightly offset to higher velocity dispersion for a given surface density from the rest of the \gls{phangs} sample.
The dispersion-surface density trends in the merger samples are more consistent with this offset \gls{phangs} trend.
There may also be evidence for the merger sample trends broadening at high surface densities, extending to higher velocity dispersions at a given surface density than the \gls{phangs} sample. This broadening in velocity dispersion would also produce a larger range of virial parameter at high surface densities (see Equation~\ref{eq:alpha_vir}).
However, the sensitivity limit in this parameter space roughly follows the upper-left edge of the contours \citep[see Figure~\num{1} from][]{Sun2020}, so that we do not have enough sensitivity to detect low surface density regions with large velocity dispersions. This effect is likely truncating the extent of the scatter in velocity dispersion seen along the trends for \ngc{} and \ngc{3256}  at lower surface densities.
To explore if the trends in the mergers truly broaden only at high surface densities or are possibly broader than the \gls{phangs} trend at all surface densities would require observations with a sensitivity a factor of two to four times better than the current dataset.

Figure~\ref{fig:dispersion_vs_surface_density_split} shows the same contours for \ngc{3256} and the \gls{phangs} galaxies, but now the sample from \ngc{} is split into pixels within \SI{1}{\kilo\parsec} of either nucleus and those beyond.
As already seen in Figures~\ref{fig:percentiles_sun_2018} and \ref{fig:percentiles_sun_2020}, pixels near the nuclei of \ngc{} exhibit higher surface densities and velocity dispersions.
The two-dimensional view also emphasizes the split between regions is not perfect as there is considerable overlap between the two sub-samples in \ngc{}.
The nuclear regions of \ngc{} are generally most consistent with the centres of (mostly barred) \gls{phangs} galaxies, while the non-nuclear region extends over nearly the complete  range limits of the full  \gls{phangs} sample (both discs and centres), likely due to the presence of the high surface density overlap region within the non-nuclear region of \ngc{}.
A similar separation of samples by nuclear and non-nuclear regions is seen in \ngc{3256} \citep[see figure~\num{7} from][]{Bru2021}.
We also note that the trend of velocity dispersion with surface density appears stronger in the non-nuclear sample from \ngc{} than the nuclear sample, such that a majority of the mass in the nuclear sample actually appears with nearly constant velocity dispersion with surface density.
This feature may indicate our choice of \gls{co}-to-H$_{2}$ conversion factor is not appropriate for regions near the nuclei of \ngc{}, and we discuss this possibility further in Section~\ref{sec:discussion}.

\section{Discussion} \label{sec:discussion}
\subsection{Comparison of the two mergers}
\ngc{} and \ngc{3256} are in different merger stages and the two systems also show significant differences in their molecular gas conditions. 
Such differences are indeed predicted by merger simulations.
Previous numerical work on replicating the morphology and kinematics of \ngc{} seems to point towards it being sometime around second pericentre passage \citep{Too1972, Bar1988, Mih1993, Kar2010, Pri2013, Ren2015}, with the large gas content suggesting that it may eventually enter the ULIRG phase \citep[e.g.,][]{Gao2001b}{}{}.
\ngc{3256} is a late-stage merger, likely entering the point of coalescence, due to its disturbed morphology and closely separated nuclei that share a common envelope \citep{Sti2013}.
In their simulations of galaxy mergers, \citet{Mor2019} show that after first pericentre passage, the \gls{sfr} of the merging galaxy relative to an isolated control galaxy increases by a factor of about five for about \SI{25}{\mega\year} before settling to a more moderate enhancement of a factor of about two.
After second pericentre passage and during coalescence, a second dramatic enhancement, with the \gls{sfr}  $\sim$\numrange{10}{30} times higher than the control galaxy, persists for about \SI{500}{\mega\year}  before more gradually decreasing.
A comparison of the  \gls{sfr} estimate of \ngc{} (ranging from \SIrange{11}{20}{\solarmass\per\year}) 
and \ngc{3256} (\SI{50}{\solarmass\per\year}) to the simulated \gls{sfr} enhancements fits their inferred merger stages. \citet{He2023} used these same simulations to investigate how the properties of the molecular clouds evolve during the merger. They find that clouds in the merger have higher virial parameters than in simulations of isolated galaxies and that the increase in virial parameter coincides with the increase in the global star formation rate.

Our observations of higher molecular gas surface densities in \ngc{} than \ngc{3256} (see e.g. Figure~\ref{fig:1d_kdes_mergers}) but lower \gls{sfr} may also fit into the evolution of dense gas in the simulations presented by \citet{Mor2019} and investigated further in \citet{He2023}.
After the first pericentre passage and until just after the second passage, the mass of cold dense gas ($n > \SI{10}{\per\cubic\centi\metre}$, $T < \SI{300}{\kelvin}$) in their fiducial merger simulation is enhanced to almost twice that in their isolated galaxy control (their figure~\num{7}).
The dense gas mass then becomes depleted relative to the isolated galaxy before rebounding to about the same mass just before coalescence, and then slowly depleting to below the mass in the isolated galaxy.
We may be catching \ngc{} close enough to its second encounter that the molecular gas mass is still maximally enhanced whereas \ngc{3256} is showing signs of the subsequent reduction of molecular gas.

To explain the higher \gls{sfr} in \ngc{3256} compared to \ngc{} despite their comparable ranges of molecular gas surface density we turn to the behaviour of the densest gas in the simulations.
While \citet{Mor2019} found the total mass of molecular gas decreased after second passage, the mass in cold ultra-dense gas ($n > \SI{1000}{\per\cubic\centi\metre}$, $T < \SI{300}{\kelvin}$) increased by a factor of about \num{100}\footnote{We note that this simulated "ultra-dense gas" is on average somewhat denser (by a factor of $\sim$~10) than the "bulk molecular gas" traced by CO observations of galaxies.}.
The fraction of cold dense gas in the ultra-dense regime increased from about \num{0.1} per cent to around \num{30} per cent, leading to the dramatic increase in the \gls{sfr} after second passage.
Thus, it could be that the total mass of very dense gas is significantly higher in \ngc{3256} than in \ngc{}, which could lead to an increase in the overall star formation rate.

While the velocity dispersions in the non-nuclear regions of \ngc{} and \ngc{3256} are quite similar, the nuclear velocity dispersions in \ngc{3256} are significantly higher than in \ngc{} (Figure~\ref{fig:percentiles_sun_2018}).
Higher velocity dispersions near the nuclei of \ngc{3256} may also be driven by the different merger stages and possibly influenced by the presence of outflows \citep{Sakamoto2013}.
Star formation is predicted to produce at most about \SI{10}{\kilo\metre\per\second} of turbulent velocity dispersion \citep{She2012, Kru2018}, with gas flows required to power larger velocity dispersions \citep{Kru2018}.
The later merger stage of \ngc{3256}, resulting in more morphological disruption and likely more significant tidal flows of gas towards the nuclei, could result in the larger velocity dispersions compared to \ngc{}.

Combining higher surface densities with similar velocity dispersions (in the non-nuclear regions) in \ngc{} as compared to \ngc{3256} results in lower virial parameters for some of the gas in \ngc{}.
While we might expect the higher \gls{sfr} in \ngc{3256} to imply lower virial parameters than in \ngc{}, we may be seeing the combination of enhanced turbulence and ultra-dense gas in \ngc{3256} making the gas appear unbound at the scales probed.
Perhaps in the more violent coalescence stages only the densest (and therefore smaller volume) portions of the molecular gas in \ngc{3256} will even approach being gravitationally bound.
In contrast, weaker tidal flows in \ngc{} \citep[cf.][]{Ueda2017}{}{} may not have transformed its \gls{ism} away from appearing largely near virial equilibrium or collapse like many of the \gls{phangs} galaxies.
The \gls{ism} in \ngc{} still being in a transitional stage may explain the wide range of virial parameters compared to \gls{phangs} and the non-nuclear regions of \ngc{3256}.

It could also be that the methods to measure \gls{sfr} observationally \citep[which measure the average \gls{sfr} over 10-100 Myr,][]{Ken2012}{}{}
 lag behind the properties of the molecular gas (which are measured "instantaneously").
The dynamical state of the molecular gas in \ngc{3256} may now be unfavorable for widespread star formation but our \gls{sfr} indicators rely on a currently existing population of stars that are the result of past gas conditions.
We must also caution that the virial parameters shown in Figures~\ref{fig:percentiles_sun_2018} through \ref{fig:percentiles_sun_2020} do not account for pressure confinement contributing to binding the observed molecular gas.
\citet{Bru2021} roughly estimate that stellar surface densities and cloud-cloud collisions between \glspl{gmc} could provide enough external pressure to balance the internal pressure for some of the gas in \ngc{3256}.

If the \gls{co} emission is optically thick in both \ngc{} and \ngc{3256} then the peak brightness temperature would be roughly equal to the kinetic temperature of the molecular gas.
The higher peak brightness temperatures we measure in \ngc{3256} would then indicate that the kinetic temperature of the gas is also higher than in \ngc{}.
Given the higher \gls{sfr} in \ngc{3256} is producing more stars to heat the molecular gas, it is not surprising that the molecular gas appears hotter in \ngc{3256} than \ngc{}.
A radiative transfer analysis of \ngc{3256} spanning several \gls{co} transitions as well as optically thin and thick isotopologues would be necessary to further investigate differences in the temperatures and densities of the molecular gas in the two mergers at \gls{gmc} scales.

As mentioned in Section~\ref{sec:samples}, the changes in the median molecular gas properties with resolution in \ngc{} are typically small relative to \gls{phangs} measured by \citet{Sun2018, Sun2020}.
Even weaker trends in median molecular gas properties were measured in \ngc{3256} by \citet{Bru2021}.
Starting with a beam that is roughly the size of a typical \gls{gmc} and then increasing the beam size, the measured surface density and peak brightness temperature would tend to decrease as a result of beam dilution.
Conversely, if the starting beam is smaller than the typical \gls{gmc} size, then increasing the beam size would likely not result in very much change in the measured surface density.
The cloud-finding analysis by \citet{Bru2022}  resulted in cloud radii slightly larger in \ngc{3256} than \gls{phangs} galaxies, pointing towards \ngc{3256} being in the second scenario described.
The weak trends measured in \ngc{} imply it is also in the second scenario and that the typical size of \glspl{gmc} may again be larger than those in \gls{phangs} galaxies.
The clouds in \ngc{} may not be as large as in \ngc{3256} since we measured slightly stronger trends in \ngc{}.
An interesting investigation would be to calculate the ``clumping factor'' defined by \citet{Ler2013} for \ngc{} and \ngc{3256} and compare to those calculated by \citet{Sun2022} for the \gls{phangs} galaxies.
If the mergers exhibit lower clumping factors than the \gls{phangs} galaxies, then a shortcut for estimating the clumping and typical \gls{gmc} size could be to calculate the changes in molecular gas property medians over a fairly modest change in resolution.

\subsection{CO-to-H$_2$ conversion factor considerations}
Since the \gls{co}-to-H$_{2}$ conversion factor influences many of the molecular gas properties presented here, we now discuss the appropriateness of our choice to use a single value (the Milky Way value) throughout \ngc{}.
A single conversion factor has the benefit that changes to the factor result in shifting the sample of quantities all together.
However, it is likely an oversimplification to assume all regions within \ngc{} truly have the same conversion factor. 
For example,  in their merger simulations, \citet{Ren2019a} find  that the \gls{co}-to-H$_{2}$ conversion factor has a significant time variation. They conclude that the conversion factor fluctuations are produced by both feedback energy and velocity dispersion.
 In their analysis of a simulation of an Antennae-like merger, \citet{Ren2019b} find  that the conversion factor has strong spatial variations as well.
 
There may be a signature in our data of the conversion factor in \ngc{} changing in the gas closer to the nuclei.
While a trend between the velocity dispersion and surface density in \ngc{} is obvious up to about \SI{1e3}{\solarmass\per\square\parsec}, for surface densities above this value there is a flattening of the trend that is most easily seen in the nuclei in Figure~\ref{fig:dispersion_vs_surface_density_split} (top panel).
Since this flattening is strongest in the nuclear sub-sample from \ngc{}, and is not seen in any of the sub-samples from \citet{Sun2020} (who adopt a radially variable conversion factor), our choice of a single conversion factor for \ngc{} may be the cause\footnote{An alternative interpretation is that the gas in the nuclei is far from self-gravitating, in which case no correlation between velocity dispersion and surface density would be expected.}.
Larger velocity dispersions and hotter gas, like the medians from \ngc{} in Figure~\ref{fig:percentiles_sun_2018} imply, can drive the conversion factor down \citep{Bol2013}.
If the conversion factor does decrease towards the nuclei in \ngc{} then the nuclear surface densities would need to be shifted to the left in Figure~\ref{fig:dispersion_vs_surface_density_split}.
An interesting consequence of this interpretation is that for pixels with surface densities roughly below about \SI{1e3}{\solarmass\per\square\parsec}, the conversion factor may not vary as much as it will in the nuclear regions.

As for the choice of the Milky Way value itself, this choice helps in simplifying comparisons with other work like \citet{Sun2018} and Galactic studies.
For example, a single conversion factor across \ngc{} and the \gls{phangs} galaxies from \citet{Sun2018} means that Figure~\ref{fig:percentiles_sun_2018} 
indicates that the surface brightness of \gls{co} emission is systematically higher in \ngc{}.
However, there is evidence that the Milky Way conversion factor may not be far off for the bulk of the \gls{co} emission in \ngc{}, even given the typical differences between spiral galaxies and starbursts/mergers \citep{Nar2011, Nar2012, Ren2019a, Ren2019b}.
\citet{Wil2003} estimated $\alpha_{\mathrm{CO(1-0)}} \SI{\approx 6.5}{\solarmass\per\square\parsec}(\si{\kelvin\kilo\metre\per\second})^{-1}$ in \ngc{} from virial mass estimates of resolved super giant molecular complexes, or a factor of $\approx$\num{1.5} larger than the Milky Way value we use here.
\citet{Sch2014} performed non-local thermodynamic equilibrium radiative transfer modeling of \ngc{} with the \textsc{radex} code on eight transitions of \gls{co} and two transitions of [CI].
To make their mass estimate consistent with that of \citet{Bra2009} they had to assume a starburst-like \gls{co} abundance and arrived at a conversion factor of $\alpha_{\mathrm{CO(1-0)}} \SI{\approx 7}{\solarmass\per\square\parsec}(\si{\kelvin\kilo\metre\per\second})^{-1}$, consistent with \citet{Wil2003}.
Scaling this conversion factor to \gls{co} (\num{2}--\num{1}), with the same $R_{21}$ we adopted, gives a value of $\alpha_{\mathrm{CO(2-1)}} \approx \SI{10}{\solarmass\per\square\parsec}(\si{\kelvin\kilo\metre\per\second})^{-1}$.
Given the observed spread in the conversion factor within starbursts of factors of $\sim$\numrange{3}{4} \citep{Bol2013}, we have opted to simplify comparisons to previous works by simply using the Milky Way value.
We will address the question of what conversion factor is appropriate for \ngc{} by combining all \gls{co} data in the present observations with additional \gls{alma} observations of several transitions of optically thin and thick \gls{co} isotopologues, at similar spatial resolution to the observations presented here (He et al. 2024, submitted).

\section{Conclusions} \label{sec:conclusions}

We have presented \gls{gmc}-scale observations of the central \SI{9}{\kilo\parsec} of \ngc{} in \gls{co} (\num{2}--\num{1}).
Maps of molecular gas mass surface density, velocity dispersion, peak brightness temperature, virial parameter, and internal turbulent pressure have been derived from these data at a range of spatial resolutions from \SIrange{55}{150}{\parsec}.
Comparisons of the pixel-by-pixel distributions of these gas properties from \ngc{} have been made at matched spatial resolution to the \gls{phangs} sample of \num{70} nearby spiral galaxies \citep{Sun2018, Sun2020}, the overlap region of \ngc{} observed in \gls{co} (\num{3}--\num{2}) \citep{Whi2014, Sun2018}, and the merger and nearest luminous infrared galaxy \ngc{3256} \citep{Bru2021}.

Relative to the \gls{phangs} galaxies, \ngc{} has some of the highest molecular gas surface densities, velocity dispersions, peak brightness temperatures, and turbulent pressures.
These gas properties measured in the discs of the \gls{phangs} galaxies are often significantly lower than those measured in \ngc{}.
The centres of the \gls{phangs} galaxies with the highest stellar masses do show some overlap with the gas properties in \ngc{}.
Virial parameters measured in \ngc{}, while spanning a large range, are much more similar to the \gls{phangs} galaxies than any other gas property.
Differences by region within \ngc{} (nuclei vs. non-nuclear) are similar to those seen within \gls{phangs} galaxies, with the most extreme gas properties near the nuclei.

Gas surface densities are similar between \ngc{} and \ngc{3256}, though the highest values are measured in \ngc{}.
On the other hand, velocity dispersions near the nuclei of \ngc{3256} and peak brightness temperatures throughout the system are significantly higher than in \ngc{}.
Higher peak brightness temperatures may be indicative of higher kinetic temperatures caused by the higher star formation rate in \ngc{3256}.
Some of the higher velocity dispersions in the nuclei of \ngc{3256} may be due to contamination from the jet in the southern nucleus and/or the outflow in the northern nucleus \citep{Sakamoto2013,Bru2021}.
However, the later merger stage of \ngc{3256} compared to \ngc{} may also be producing more intense merger-driven flows into the nucleus of \ngc{3256} that could produce more turbulence and result in larger line widths.
The wider but systematically lower range of virial parameters in \ngc{} compared to \ngc{3256} could also be a consequence of the differing merger stages.
\ngc{3256} may be leaving the starburst phase as coalescence of the progenitor galaxies increases the turbulence and suppresses further star formation, but \ngc{} may be caught in the transition between bursts of enhanced star formation.

Next steps will focus on comparisons of several \gls{co} isotopologues and transitions across the entire region observed in this work (He et al. 2024, submitted).
$J$=\num{2}--\num{1} transitions of $^{13}$\gls{co} and C$^{18}$O as well as the $J$=\num{1}--\num{0} transition of $^{12}$\gls{co} were also observed in this project and will allow for exact matching of $uv$ coverage as well as spatial and spectral resolutions.
Additional observations have been obtained with \gls{alma}, mapping the same field of view with the same array combinations in $^{12}$\gls{co} (\num{3}--\num{2}) and $^{13}$\gls{co} (\num{1}--\num{0}), extending the area probed by these transitions at cloud scales.
With three transitions of $^{12}$\gls{co} and two transitions of $^{13}$\gls{co} a radiative transfer analysis can constrain the densities and temperatures of the molecular gas in \ngc{}, and help to estimate the \gls{co}-to-H$_{2}$ conversion factor at cloud scales (He et al. 2024, submitted).

In summary, these two nearby galaxy mergers show significantly higher molecular gas surface densities, velocity dispersions, peak brightness temperatures, and turbulent pressures on giant molecular cloud scales compared to spiral galaxies from the \gls{phangs} survey. However, there are also significant differences between the two mergers that may be linked to the phase of the merger. For example, the larger velocity dispersion and higher gas temperature of \ngc{3256} may be linked to enhanced feedback from star formation in its more advanced merger phase compared to \ngc{}. In contrast, there may be a current up-tick in the star formation rate in \ngc{}, highlighting time variability as an important factor in understanding galaxy mergers observationally as well as theoretically \citep{Mor2019,He2023}. Additional studies on molecular cloud scales of nearby starburst mergers as well as the pre- and post-merger stages will be important for placing these results in a wider context and ultimately informing our understanding of the high star formation rates seen in galaxies at high redshift.

\section*{Acknowledgements}

We thank the referee for comments that improved the presentation and discussion of the paper. The research of NB and HH is partially supported by grants from the Natural Sciences and Engineering Research Council of Canada through the New Technologies for Canadian Observatories program.
The research of CDW and ER is supported by grants from the Natural Sciences and Engineering Research Council of Canada (NSERC), and also for CDW by the Canada Research Chairs program. JS acknowledges support by
NSERC through a Canadian Institute for Theoretical
Astrophysics (CITA) National Fellowship. FB
acknowledges funding from the European Research Council
(ERC) under the European Unions Horizon 2020 research and
innovation program (grant agreement No.726384/Empire). ES
acknowledges funding from the ERC under the European Unions Horizon 2020 research and
innovation program (grant agreement No. 694343).

We acknowledge the use of the ARCADE (ALMA Reduction in the CANFAR Data Environment) science platform.
ARCADE is a ALMA Cycle 7 development study with support from the National Radio Astronomy Observatory, the North American ALMA Science Centre, and the National Research Centre of Canada.

This research made use of \textsc{Astropy}, a community-developed core \textsc{Python} package for Astronomy \citep[\url{http://www.astropy.org},][]{astropy2013,astropy2018}.
This research also made use of the \textsc{SciPy} \citep{Vir2020}, \textsc{Matplotlib} \citep{Hun2007}, \textsc{NumPy} \citep{van2011}, \textsc{pandas} \citep{McK2010}, \textsc{Jupyter Notebook} \citep{Klu2016}, and \textsc{spectral-cube} \citep{Gin2019} \textsc{Python} packages.
This research has made use of the \textsc{R} programming language \citep{R2019}.
This research has made use of the \gls{carta} \citep{Com2021}.
This research has made use of NASA’s Astrophysics Data System.
This research has made use of the VizieR catalogue access tool \citep{Och2000}.
This research has made use of the NASA/IPAC Extragalactic Database (NED), which is funded by the National Aeronautics and Space Administration and operated by the California Institute of Technology.
This research has made use of the SIMBAD database, operated at CDS, Strasbourg, France \citep{Wen2000}.

\section*{Data Availability}
This paper makes use of the following ALMA data: ADS/JAO.ALMA\#2018.1.00272.S (accessed from the \gls{alma} Science portal at \url{almascience.org}).
ALMA is a partnership of ESO (representing its member states), NSF (USA) and NINS (Japan), together with NRC (Canada), MOST and ASIAA (Taiwan), and KASI (Republic of Korea), in cooperation with the Republic of Chile.
The Joint ALMA Observatory is operated by ESO, AUI/NRAO and NAOJ.
The National Radio Astronomy Observatory is a facility of the National Science Foundation operated under cooperative agreement by Associated Universities, Inc.

The primary-beam cube from project ADS/JAO.ALMA\#2011.0.00876.S was retrieved from the JVO portal (\url{http://jvo.nao.ac.jp/portal}) operated by the NAOJ.

The derived data generated in this research will be shared on reasonable request to the corresponding author.



\bibliographystyle{mnras}
\bibliography{references}



\appendix

\renewcommand{\thesubsection}{\Alph{section}}
\setcounter{section}{0}

\section{Modifications to the PHANGS-ALMA interferometric imaging pipeline} \label{sec:pipeline_modifications_appendix}

Starting from commit 5ef53d3, we made modifications to the PHANGS-ALMA imaging pipeline to handle the additional array configurations and spectral lines present in the \ngc{} observations compared to the bulk of the \gls{phangs} observations and to address some code bugs.
The first change was to automate the continuum subtraction step based on the user-defined spectral line centre and width parameters.
Since the \gls{phangs} observations consisted of only one main array configuration combined with the \gls{aca} and \gls{tp}, the ability to combine two main array configurations had to be added.
This modification also included adding angular scales to use in Band \num{6} multi-scale cleaning of our extended main array configuration data.
Additionally, \gls{phangs} data consisted of only Band 6 observations, primarily focused on $^{12}$\gls{co} ($J$=\num{2}--\num{1}).
Steps like channel binning had to be made to handle additional lines in the \ngc{} observations, e.g. $^{12}$\gls{co} (\num{1}--\num{0}), $^{13}$\gls{co} (\num{2}--\num{1}), \gls{cn} (\num{1}--\num{0}), etc.
This modification also included the addition of angular scale specifications for the Band \num{3} multi-scale cleaning.
Coding bugs were fixed that prevented clean masks from being used in the single-scale clean step and from manually specifying image and pixel sizes. At this stage we also added the ability to manually specify the ``robust,'' ``nchan,'' ``start,'' ``uvrange,'' and ``uvtaper'' arguments to \textit{tclean}.

Given the volume of data and computing resources available, we had to switch to setting ``chanchunks'' to \num{-1} in \textit{tclean} since our highest resolution cubes could not fit in memory all at once.
To further improve memory management and processing speed, we removed additional image padding in right ascension and declination 
during imaging and instead added padding of the interferometric cubes before feathering to ensure the angular coverage was at least as large as the \gls{tp} cubes.
The final change related to the cube size was adding a check during writing masks to disk that would do the writing in batches of channels to avoid memory issues for our largest cubes that were about \SI{16}{\giga\byte} in size.
Since the pre-imaging regridding would not guarantee exactly the same spectral channels would be filled with data across all arrays, we added a call to the \textit{split} task to remove one channel at each end of the imaging \glspl{spw} to avoid blank channels being included in the cube for cleaning.
Finally, substantial effort was also spent on properly capturing all \gls{casa} logging output and redirecting it to output files.
This message capturing was missing in version one and made it difficult to ensure each step was performed correctly as well as verifying adjustments to the code for the Antennae observations.

\begin{figure*}
    \centering
    \includegraphics{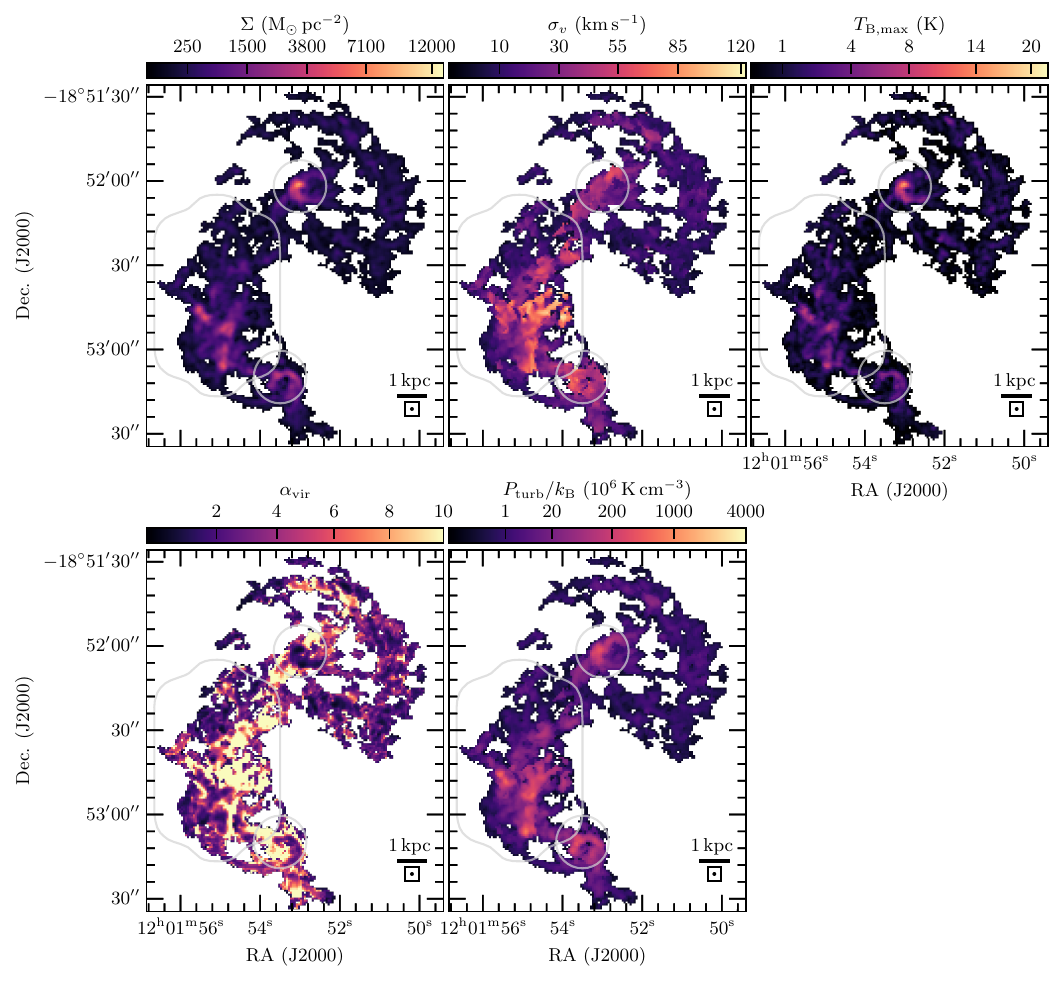}
    \caption{
        The same as Figure~\ref{fig:maps_55} but showing the molecular gas properties at \SI{150}{\parsec} resolution.
        See the Figure~\ref{fig:maps_55} caption for more details.
    }
    \label{fig:maps_150}
\end{figure*}

Table~\ref{tab:observations} gives a summary of the observations and the calibration method used for each individual observation. Figure~\ref{fig:maps_150} show the resulting maps processed at  \SIlist{150}{\parsec} resolution for comparison to Figure~\ref{fig:maps_55} at  \SIlist{55}{\parsec} resolution.

\begin{table*}
    \caption{Summary of observations and calibration methods.}
    \begin{threeparttable}
        \label{tab:observations}
        \sisetup{table-number-alignment=center,table-sign-mantissa}
        \begin{tabular}{
            @{}
            l
            S[table-format=1.0]
            c
            S[table-format=1.0]
            S[table-format=2.0]
            S[table-format=4.0]
            c
            c
            @{}
        }
            \toprule
             Array           & {Band} & Observation & {Repeats} & {Minimum}           & {Maximum}           & \acrshort{casa} & Pipeline \\
                             &        & date        &           & {baseline\tnote{a}} & {baseline\tnote{a}} & version         & version \\
                             &        &             &           & {(m)}               & {(m)} \\
             \midrule
             \SI{12}{\metre} & 3      & 2018 Oct 30 & 1         & 15                  & 1400                & 5.4.0-68        & 42030M\tnote{b} \\ 
             \SI{12}{\metre} & 3      & 2018 Nov 1  & 1         & 15                  & 1400                & 5.4.0-68        & 42030M\tnote{b} \\ 
             \SI{12}{\metre} & 3      & 2018 Nov 3  & 1         & 15                  & 1400                & 5.4.0-68        & 42030M\tnote{b} \\ 
             \SI{12}{\metre} & 3      & 2018 Nov 4  & 2         & 15                  & 1400                & 5.4.0-68        & 42030M\tnote{b} \\ 
             \SI{12}{\metre} & 3      & 2018 Nov 6  & 2         & 15                  & 1400                & 5.4.0-68        & 42030M\tnote{b} \\ 
             \SI{12}{\metre} & 3      & 2018 Nov 8  & 1         & 15                  & 1400                & 5.4.0-68        & 42030M\tnote{b} \\ 
             \SI{12}{\metre} & 3      & 2018 Nov 14 & 1         & 15                  & 1400                & 5.4.0-68        & 42030M\tnote{b} \\ 
             \SI{12}{\metre} & 3      & 2018 Nov 17 & 1         & 15                  & 1400                & 5.4.0-68        & 42030M\tnote{b} \\ 
             \SI{12}{\metre} & 6      & 2018 Nov 11 & 2         & 15                  & 1400                & 5.4.0-70        & 42254M\tnote{b} \\ 
             \SI{12}{\metre} & 6      & 2019 Apr 16 & 2         & 15                  & 741                 & 5.4.0-70        & 42254M\tnote{b} \\ 
             \SI{12}{\metre} & 3      & 2019 Jan 1  & 2         & 15                  & 500                 & 5.4.0-70        & 42254M\tnote{b} \\ 
             \SI{12}{\metre} & 3      & 2019 Jan 5  & 1         & 15                  & 500                 & 5.4.0-70        & 42254M\tnote{b} \\ 
             \SI{12}{\metre} & 6      & 2019 Jan 11 & 1         & 15                  & 314                 & 5.4.0-70        & 42254M\tnote{b} \\ 
             \SI{7}{\metre}  & 3      & 2018 Oct 23 & 1         & 9                   & 45                  & 5.4.0-68        & 42030M\tnote{b} \\
             \SI{7}{\metre}  & 3      & 2018 Nov 6  & 2         & 9                   & 49                  & 5.4.0-68        & 42030M\tnote{b} \\
             \SI{7}{\metre}  & 3      & 2018 Dec 4  & 1         & 9                   & 49                  & 5.4.0-68        & 42030M\tnote{b} \\
             \SI{7}{\metre}  & 3      & 2018 Dec 9  & 1         & 10                  & 49                  & 5.4.0-68        & 42030M\tnote{b} \\
             \SI{7}{\metre}  & 3      & 2018 Dec 11 & 1         & 9                   & 49                  & 5.4.0-68        & 42030M\tnote{b} \\
             \SI{7}{\metre}  & 3      & 2018 Dec 15 & 1         & 9                   & 49                  & 5.4.0-68        & 42030M\tnote{b} \\
             \SI{7}{\metre}  & 3      & 2018 Dec 16 & 2         & 9                   & 49                  & 5.4.0-68        & 42030M\tnote{b} \\
             \SI{7}{\metre}  & 3      & 2018 Dec 17 & 1         & 9                   & 49                  & 5.4.0-68        & 42030M\tnote{b} \\
             \SI{7}{\metre}  & 6      & 2018 Oct 16 & 4         & 9                   & 49                  & 5.4.0-68        & 42030M\tnote{b} \\
             \SI{7}{\metre}  & 6      & 2018 Oct 20 & 2         & 9                   & 49                  & 5.4.0-68        & 42030M\tnote{b} \\
             \acrshort{tp}   & 3      & 2018 Oct 23 & 1         & {\ldots}            & {\ldots}            & 4.7.2           & ca9f82c\tnote{c} \\
             \acrshort{tp}   & 3      & 2018 Nov 12 & 1         & {\ldots}            & {\ldots}            & 4.7.2           & ca9f82c\tnote{c} \\
             \acrshort{tp}   & 3      & 2018 Nov 25 & 1         & {\ldots}            & {\ldots}            & 4.7.2           & ca9f82c\tnote{c} \\
             \acrshort{tp}   & 3      & 2018 Nov 27 & 2         & {\ldots}            & {\ldots}            & 4.7.2           & ca9f82c\tnote{c} \\
             \acrshort{tp}   & 3      & 2018 Nov 29 & 1         & {\ldots}            & {\ldots}            & 4.7.2           & ca9f82c\tnote{c} \\
             \acrshort{tp}   & 3      & 2018 Dec 3  & 2         & {\ldots}            & {\ldots}            & 4.7.2           & ca9f82c\tnote{c} \\
             \acrshort{tp}   & 3      & 2018 Dec 4  & 1         & {\ldots}            & {\ldots}            & 4.7.2           & ca9f82c\tnote{c} \\
             \acrshort{tp}   & 3      & 2018 Dec 5  & 1         & {\ldots}            & {\ldots}            & 4.7.2           & ca9f82c\tnote{c} \\
             \acrshort{tp}   & 3      & 2018 Dec 6  & 1         & {\ldots}            & {\ldots}            & 4.7.2           & ca9f82c\tnote{c} \\
             \acrshort{tp}   & 3      & 2018 Dec 8  & 2         & {\ldots}            & {\ldots}            & 4.7.2           & ca9f82c\tnote{c} \\
             \acrshort{tp}   & 3      & 2018 Dec 9  & 3         & {\ldots}            & {\ldots}            & 4.7.2           & ca9f82c\tnote{c} \\
             \acrshort{tp}   & 6      & 2018 Oct 24 & 1         & {\ldots}            & {\ldots}            & 4.7.2           & ca9f82c\tnote{c} \\
             \acrshort{tp}   & 6      & 2018 Oct 30 & 1         & {\ldots}            & {\ldots}            & 4.7.2           & ca9f82c\tnote{c} \\
             \acrshort{tp}   & 6      & 2018 Oct 31 & 2         & {\ldots}            & {\ldots}            & 4.7.2           & ca9f82c\tnote{c} \\
             \bottomrule
         \end{tabular}
         \begin{tablenotes}
             \item [] \emph{Notes.} All interferometric observations were multiple-pointing mosaics covering an area approximately $\num{2.8} \times \SI{1.9}{\arcmin}$, centred on 
             ($12^{\mathrm{h}}01^{\mathrm{m}}53^{\mathrm{s}}$, 
             \ang{-18;52;30}, J2000).
             \item[a] Projected for source position on the sky.
             \item[b] \gls{alma} observatory pipeline. All observatory pipeline versions were followed by ``(Pipeline-CASA54-P1-B).''
             \item[c] \gls{phangs} \gls{tp} pipeline obtained from \url{https://github.com/PhangsTeam/TP_ALMA_data_reduction}. Version hash is the \textsc{Git} commit at which we obtained the pipeline.
         \end{tablenotes}
    \end{threeparttable}
\end{table*}

\section{Data tables} \label{sec:data_tables}
\begin{table*}
	\centering
	\caption{Binned data for NGC~4038/9 used in this paper}
	\label{table:antennae_pixel_data}
	\begin{tabular}{lccccccccccccc} 
		\hline
Beam & R.A. & Decl. 
& $\log \Sigma_{\rm mol}$ & $\sigma_{\rm v}$ 
& T$_{\rm B}$ & $\alpha{\rm vir}$ & $\log{\rm P_{\rm turb}}$
& NGC4908 & NGC4039 & overlap
\\
FWHM\footnotemark[1] &&&&&&&& distance & distance & region? \\
		\hline
55.0 & 180.470384 & -18.892159  & 1.684  & 3.3 & 0.91 & 1.91 & 4.677 & 9.587 & 2.518 & False \\
80.0 & 180.470997 & -18.892576  & 1.646  & 3.5 & 0.82 & 1.60 &  4.524	&	9.746 & 2.599 & False \\
90.0 & 180.470997 &  -18.892589 & 1.6450   & 3.6 &  0.73 & 1.61 & 4.464 & 9.751  & 2.604  & False \\	
120.0 & 180.471988 &  -18.892628 & 1.528 & 4.1 & 0.50 & 1.98 & 4.378 & 9.775 & 2.546 & False 	 \\
150.0 & 180.471988 & -18.892589 & 1.420 & 3.4 & 0.49 & 1.36  & 4.003 & 9.760 & 2.531 & False 	 \\
	\hline

	\end{tabular}
\\
\footnotemark[1] This table is available in its entirety in machine-readable form. \\
\end{table*}

\begin{table*}
	\centering
	\caption{Binned data for NGC~3256 from \citet{Bru2021} used in this paper}
	\label{table:ngc3256_pixel_data}
	\begin{tabular}{lccccccccccccc} 
		\hline
Beam & R.A. & Decl. 
& $\log \Sigma_{\rm mol}$ & $\sigma_{\rm v}$ 
& T$_{\rm B}$ & $\alpha{\rm vir}$ & $\log{\rm P_{\rm turb}}$
& N nucleus & S nucleus & north & south & west 
\\
FWHM\footnotemark[1] &&&&&&&& distance & distance & jet? & jet? & region? \\
		\hline
55.0 & 156.958656 & -43.908402 & 1.459  & 8.1 & 1.09 & 19.13 & 5.226  & 23.257 & 92.168 & False & False & True \\
80.0 & 156.958290 & -43.908435 &  1.398  & 7.8 & 0.79 & 14.07  & 4.971  
& 24.641  & 95.893  & False & False & True  \\
120.0 & 156.959157 &  -43.909034 &  1.348  & 7.1 & 0.69 & 8.77 & 4.665 
& 26.925  & 131.593  & False & False & True  \\
 	\hline
	\end{tabular}
\\
\footnotemark[1] This table is available in its entirety in machine-readable form. \\
\end{table*}

We provide the binned data used in this paper for both \ngc{} and \ngc{3256} as two separate machine-readable tables (see Table~\ref{table:antennae_pixel_data} and~\ref{table:ngc3256_pixel_data} for format). In these tables, each row reports our measurements for one pixel in one galaxy; pixel values at different physical resolutions are reported sequentially. The contents of the rows are as follows:

\begin{enumerate}
    \item The resolution (beam FWHM) of the dataset in parsecs. 
    \item the right ascension and declination of the pixel in degrees.
    \item The logarithmic value of the molecular gas surface density, $\Sigma_{\rm mol}$ in M$_\odot$ pc$^2$ ; the velocity dispersion, $\sigma_{\rm v}$ in km s$^{-1}$; the peak brightness temperature, $T_{\rm B}$ in K; the virial parameter, $\alpha_{\rm vir}$; and the logarithmic value of the turbulent pressure, $P_{\rm turb}$, in K km s$^{-1}$.
    \item The projected distances of the pixel from each of the two nuclei, in kpc.
\end{enumerate}

Table~\ref{table:antennae_pixel_data} also indicates whether or not a pixel lies in the overlap region. Table~\ref{table:ngc3256_pixel_data} indicates whether or not a pixel lies in the north or south portion of the jet or the west exclusion zone; see \citet{Bru2021} for details.


\bsp	
\label{lastpage}
\end{document}